\documentclass[pra,aps]{revtex4}
\usepackage{graphics}
\date{November 2, 2010 }

\begin{document}
\title{Integrability breakdown in longitudinaly trapped, one-dimensional bosonic gases}
\author{Igor E. Mazets
}                     
%
%
\affiliation{Institute of Atomic and Subatomic Physics, Vienna University of Technology, 
1020 Vienna, Austria;   \\ 
Ioffe Physico-Technical Institute, 194021 St.Petersburg, Russia }
%

%
\abstract{
A system of identical bosons with short-range (contact) interactions is studied. Their  
motion is confined to one dimension by a tight 
lateral trapping potential and, additionally, subject to a weak harmonic confinement in the longitudinal 
direction. Finite delay time associated with penetration of quantum particles through each other 
in the course of a pairwise one-dimensional collision in the presence of the longitudinal potential 
makes the system non-integrable and, hence, provides a mechanism for relaxation to thermal equilibrium. 
To analyse this effect quantitatively in the limit of a non-degenerate gas, we develop a system of kinetic 
equations and solve it for small-amplitude monopole oscillations of the gas. The obtained damping rate 
is long enough to be neglected in a realistic cold-atom experiment, and therefore longitudinal trapping 
does not hinder integrable dynamics of atomic gases in the 1D regime. 
%
} 
} 
\maketitle
\section{Introduction}
\label{intro}
One-dimensional systems (1D) \cite{Popov,Giamarchi} often provide suitable models to study the 
fundamental processes of quantum dynamics, coherence and noise in interacting many-body systems. 
Integrability is one of such fundamental questions most closely related to the 1D physics.
In an integrable system \cite{Thacker,Yurovsky1}, the number of integrals of motion is exactly 
equal to the number of degrees of freedom. Thus the observables, which can be measured experimentally, 
namely, single-particle (momentum and position distribution) or few-particle (correlation functions) 
ones retain the information on the initial conditions forever. As a result, a quenched (i.e. suddenly 
driven off the equilibrium) integrable system may undergo relaxation towards the generalized 
Gibbs (or fully constrained thermodynamics) ensemble \cite{Rigol1}, instead of the thermal equilibrium 
state. A similar type of relaxation may be observed even in non-integrable systems \cite{Altman1}. 
Also local quantities of the Bose-Hubbard model are shown \cite{Fl1} to approach the thermal Gibbs 
state much faster than non-local ones, such as the density-density correlation function. 
For a very wide class of non-integrable systems, however, the eigenstate thermalization hypothesis 
\cite{ethh} (see also \cite{Rigol2}) 
may hold, enabling fast relaxation to thermal equilibrium. Of corse, the term 
``thermalization"  with regard to closed (isolated from environment) systems means that all 
experimentally measurable observables approach their thermal distributions; the information on the 
initial conditions becomes contained in many-particle correlations (ultimately, in correlations 
involving all particles in the system), whose measurement is not feasible in practice.  

An uniform 1D system of indistinguishable bosons interacting with each other 
via pairwise delta-functional potential is known to be integrable and described by the Lieb-Liniger 
model \cite{LL1,LL2}. It can be, up to certain extent, implemented in an ultracold-atom experiment  
in optical lattices \cite{WeissTG,Morsch} or on atom chips \cite{Folman,Reichel,Ketterle,chiprev}, 
provided that both the temperature and mean interaction energy per atom are well below the excitation 
energy of the first excited state of the radial trapping Hamiltonian. Under these conditions, the 
radial motion of atoms is confined to the ground state of the radial trapping Hamiltonian. 
Ultracold atomic systems 
deeply in the 1D regime can be prepared in optical lattices \cite{WeissNC} and on atom chips \cite{Ho1}. 
However, in reality no system is perfectly 1D, but the actual question is, on which timescale it can be 
described as 1D. For ultracold atoms under tight lateral confinement one-dimensionality and, 
hence, integrability are lifted by atomic interactions causing virtual population of excited 
radial modes, even if the energy of thermal motion is too low, and thermal excitation of the radial modes 
is strongly suppressed by a Boltzmannian exponential factor. The role of the virtual radial excitations in 
the dynamics of ultracold atomic gases in tight waveguides has been first noticed in the context of 
macroscopic flow of degenerate atomic gas through a waveguide \cite{SPR} and decay of mean-field solitons 
\cite{Muryshev}. Microscopically, virtual radial excitations give rise to effective 3-body elastic collisions,  
which have been suggested as the source of thermalization 
in ultracold atomic gases on atom chips \cite{Mazets1} (see also Refs. 
\cite{Mazets2,Glazman}). Interestingly, 
if the interatomic repulsion is so strong that the system approaches the Tonks-Girardeau limit, 
interatomic correlations reduce the rate of thermalizing collisions \cite{Mazets2,Glazman,Mazets3}. The 
latter conclusion is in agreement with the results  of the quantum ``Newton craddle" experiment \cite{WeissNC}. 

However, this is not the only source of non-integrability. An ultacold atomc cloud in a real 
experiment is also confined longitudinally, although the ratio of the fundamental frequencies of the longitudinal 
($\omega _\Vert $) and transversal ($\omega _\perp $) trapping may be as small as 
$\omega _\Vert /\omega _\perp \sim 10^{-3}$. Obviously, an ideal gas 
in 1D in the presence of harmonic confinement is an integrable system, and so is the gas of impenetrable bosons 
\cite{Minguzzi}. However, the longitudinal confinement should lift the integrability of the Lieb-Liniger model 
for any finite atomic interaction strength. Estimating the corresponding relaxation rate is the subject 
of the present paper. 

\section{Physical model}
\label{sec:1}

The Hamiltonian of the system of $N$ identical bosons of mass $m$ confined to 1D and  also 
harmonically trapped in the longitudinal direction  reads as 
\begin{equation} 
{\hat H}=\sum _{j=1}^N \left[ -\frac {\hbar ^2}{2m} \frac {\partial ^2}{\partial x_j^2} +
\frac { m\omega _\Vert ^2}2 x_j^2 +
\frac {\hbar ^2 c}{m}\sum _{i=j+1}^N \delta (x_j -x_i) \right] 
. \label{k:1.1} 
\end{equation} 
Here $c$ is the pairwise interaction strength. In the limit $\omega _\Vert \rightarrow 0$ 
we recover the Hamiltonian of the Lieb-Liniger model \cite{LL1,LL2}. 
Virtual excitations of the radial modes \cite{SPR,Muryshev,Mazets1,Mazets2,Glazman} are not 
included into the model Eq. (\ref{k:1.1}), therefore we can investigate the integrability breakdown 
solely due to the longitudinal confinement. The latter is assumed to be harmonic in order to 
distinguish the effect under study from dephasing of the atomic motion occuring in an 
anharmonic trap.  

In what follows we consider a non-degenerate, weakly interacting 1D bosonic gas. 
The non-degeneracy means that the atomic motion can be 
viewed as quasi-classical propagation in the longitudinal direction 
of distinct wave packets localized of a size much smaller than the mean 
interatomic distance. However, as we noted in Sec.~\ref{intro}, the radial motion of atoms remains 
quantized, confined to the ground state of the radial trapping Hamiltonian. This is exactly the 
ponit, where our models differs from previous calculations of damping of excitations in a 
classical atomic gas in very elongated, but still three-dimensional (3D) traps \cite{Stringari1}. 
In the latter case, (i) time delay in the 
process of scattering is neglected and (ii) collisions are 
essentially 3D and lead to scattering to a whole manifold of radial modes.  

A collision of two atoms is not an instantaneous event, but entails a time 
delay in propagation of the two colliding wave packets. The delay time is \cite{tdel} 
\begin{equation} 
\tau = \frac 1{u} \frac {\partial \varphi }{\partial k_r}  , 
\label{k:1.2} 
\end{equation} 
where $u$ is the relative velocity of the collision (we assume here wave packet propagation in 1D with $u>0$), 
$\hbar k_r =mu/2$ is the corresponding momentum 
($m/2$ is the reduced mass) and $\varphi $ is the phase shift of the transmitted wave after 1D scattering 
on the given pairwise interaction potential. Having the formula $\varphi =-\arctan [c/(2k_r)]$ for 
scattering on the contact (delta-functional) potential \cite{LL1}, it is easy to obtain  
\begin{equation} 
\tau = {mc}/[\hbar k_r(4k_r^2+c^2)] .
\label{k:1.2a} 
\end{equation}      
Since in a weakly-interacting gas $k_r\gg c$ for almost every 
collision, 
\begin{equation}
\tau \approx mc/(4\hbar k_r^3)= 2\hbar ^2 c/(m^2u^3) . 
\label{k:1.3} 
\end{equation} 
Formally, Eq. (\ref{k:1.3}) coincides with the classical expression $\tau =[2/(mu^3)] 
\int _{-\infty }^{+\infty } dx\, U_{12}(x) $ for the delay time of a 
fast particle of mass $m/2$ traversing the potential $U_{12}(x)$ [``fast" meaning that the particle kinetic 
energy $mu^2/4$ at $x\rightarrow \pm \infty $ is much larger than the maximum value of $U_{12}(x)$]. 
Of course, in a case of the repulsive ($c>0$) delta-functional potential the classical description fails, 
but Eq. (\ref{k:1.3}) holds also in the regime of quantum tunneling through a weak potential barrier. 
If the atomic interaction is attractive, we formally obtain $\tau <0$. However, the theoretical 
approach developed in Sec.~\ref{kin:eqs} assumes $\tau >0$. Therefore we consider in this paper only gases 
with repulsive interactions ($c>0$).  

The finite delay time does not change the statistical distribution of atomic velocities in an infinite 
waveguide in the thermodynamic limit. However, the situation becomes different for the trapped gases. 
The relative motion of two particles is assumed to be classical in the intervals between collisions, 
and the center-of-mass motion is always treated classically. 
Assume that at time $t_0$ two particles with the velocities $v_1(t_0)\equiv v_{c1}$ and 
$v_2(t_0)\equiv v_{c2}$ collide at the point 
$x_1(t)=x_2(t)\equiv x_c$. We assume the following model: during the collision time $\tau $, which depends 
on the absolute value of the relative velocity $|v_{c1}-v_{c2}|$, the two atoms form a composite particle of 
mass $2m$ moving in the potential $m\omega _\Vert ^2X^2$, $X$ being the coordinate of the composite particle,   
with the initial conditions $X(t_0) =x_c$, $V(t_0)\equiv 
\dot{X}(t_0)=(v_{c1}+v_{c2})/2$. The kinetic energy of the 
relative motion is ascribed to an ``internal degree of freedom"  of the compound particle, thus ensuring 
energy conservation. After the time $\tau $ the compaund particle disintegrates, releasing two atoms 
at the point 
\begin{equation} 
x_{1,2}(t_0+\tau )=x_c\cos \omega _\Vert \tau +\frac {v_{c1}+v_{c2}}{2\omega _\Vert }\sin \omega _\Vert \tau .
\label{k:1.4} 
\end{equation} 
The velocities of the particles are determined by the center-of-mass velocity and the relative-motion kinetic 
energy stored internally in the compound particle and  released back at $t=t_0+\tau $: 
\begin{equation} 
v_{1,2}(t_0+\tau )=-\omega x_c\sin \omega _\Vert \tau +\frac {v_{c1}+v_{c2}}{2}\cos \omega _\Vert \tau 
\pm \frac {v_{c1}-v_{c2}}2 .
\label{k:1.5} 
\end{equation}
Since the atoms are identical, 
there is no one-to-one correspondence between the upper (lower) sign in front of the last term and the index (1 or 2) 
numbering the atoms. Although 
the total energy of the pair of the atoms is conserved exactly, the single-atom energies are not. They will be 
shifted by $\mp \frac m2(v_{c1}-v_{c2})[ \frac 12 (v_{c1}+v_{c2})(1-\cos \omega _\Vert \tau )+ 
\omega _\Vert x_c \sin \omega _\Vert \tau ] $. 

\section{Kinetic equations} 
\label{kin:eqs}
To quantify the single-atom energy redistribution noticed in the previous section, we derive a set of 
kinetic equations describing 1D atomic collisions. We assume a two-component 
model: the system consists of single atoms chrarcterized by a joint co-ordinate and velocity distribution 
function $f(x,v,t)$ and diatomic compound particles representing atomic pairs, not actually bound but 
remaining close to each other during the delay time $\tau $, which is considered as the effective 
collision duration. The distribution function of compound pparticles $F(x,V,u,t)$ has, apart from the time 
$t$, three arguments: $x$ stands here for the compound 
particle co-ordinate (center of mass of two colliding atoms), 
$V$ is  the center-of-mass velocity, and $u$ is the relative velocity of two atoms 
before the collision; $u$ can be considered also as a label for ``internal excitations" of a compound 
particle. Because of idistinguishability of the atoms, it is sufficient to consider only positive values of 
$u$. We assume  that the gas is dilute, i.e.,  only a small fraction of the atomic 
ensemble is simultaneously contained in compound particles. 
A rate of 1D collisions experienced by an atom with the velocity $v$ is 
$\int _{-\infty }^\infty dv^\prime \, |v-v^\prime | \, f(x,v,t)f(x,v^\prime ,t)$, and each collision is 
followed by compound particle formation (it is impossible for two identical, interacting bosons to 
tunnel through each other in 1D without delay in propagation).

Also we assume that a compound particle does not 
experience collisions with other atoms during its lifetime. To make the set of kinetic equations simple, 
we assume (instead of a well-defined (deterministic) time interval between formation and disintegration 
of a compound particle) 
that a compound particle undergoes exponential decay with the rate $1/\tau _u$. The lower index 
index $u$ explicitly indicates the dependence (\ref{k:1.2a}) of the delay time on the relative velocity  
$u$ of the atoms before collisions. The products of the decay are two atoms emerging at the same point 
with the velocities $V\pm u/2$. The equality of the relative velocities before and after the collision is 
specific for motion in a harmonic potential, where the center-of-mass and relative-motion degrees of 
freedom are decoupled. 

The corresponding set of kinetic equations now can be easily written:  
\begin{eqnarray} 
\frac \partial {\partial t} f(x,v,t)+v\frac \partial {\partial x} f(x,v,t)- 
\omega _\Vert ^2 x \frac \partial {\partial v} f(x,v,t)=&& \nonumber  \\ 
-\int _{-\infty }^\infty dv^\prime \, {\bigg [} |v-v^\prime | \, f(x,v,t)f(x,v^\prime ,t)-\qquad && \nonumber \\ 
\frac 1{\tau _{|v-v^\prime |}}F\left(x,
\frac {v+v^\prime }2, |v-v^\prime | ,t\right)  {\bigg ]} , && \label{k:2.1a} \\ 
\frac \partial {\partial t} F(x,V,u,t)+V\frac \partial {\partial x} F(x,V,u,t)- \qquad \qquad 
&& \nonumber  \\ 
\omega _\Vert ^2 x \frac \partial {\partial V} F(x,V,u,t)=-\frac 1{\tau _u }F(x,V,u,t)+&& \nonumber  \\
u\, f\left( x,V+\frac u2,t\right) f\left( x,V-\frac u2,t\right) . &&
\label{k:2.1b} 
\end{eqnarray}  
It belongs to a broad family of equation sets describing 
chemical reactions on the kinetic level \cite{flow}. It conserves the total number of atoms 
\begin{equation} 
N=\int _{-\infty }^\infty dx \int _{-\infty }^\infty dv \left[ f(x,v,t) +2\int _{0}^\infty du\, F(x,v,u)\right] 
\label{k:2.2} 
\end{equation} 
and the total energy of the system 
\begin{eqnarray} 
E&=&\int _{-\infty }^\infty dx \int _{-\infty }^\infty dv \, \varepsilon _{a}f(x,v,t) + \nonumber \\ && 
\int _{-\infty }^\infty dx \int _{-\infty }^\infty dV \int _{0}^\infty du\, 
\varepsilon_{cp} F(x,V,u,t)  ,  
\label{k:2.3} 
\end{eqnarray} 
where $\varepsilon _a =m(\omega _\Vert ^2x^2+v^2)/2$ and $\varepsilon _{cp} =m(\omega _\Vert ^2x^2+V^2+u^2/4)$ 
are the single-atom and compound-particle energies, respectively. 
Eqs. (\ref{k:2.1a},~\ref{k:2.1b}) are also consistent with 
Newton's second law $d^2 X_N/dt^2=-\omega _\Vert ^2 X_N$ for the center of mass 
$X_N=\int _{-\infty}^\infty dx\, x \int _{-\infty}^\infty dv \, [ f(x,v,t)+2\int _{0}^\infty du\, 
F(x,v,u,t)] /N$ of the $N$-atom system. 

Eliminating $F(x,V,u,t)$, we arrive, by an exact transformation, to an equation for the distribution 
functions of atoms only: 
\begin{eqnarray} 
\frac \partial {\partial t} f(x,v,t)+v\frac \partial {\partial x} f(x,v,t)- 
\omega _\Vert ^2 x \frac \partial {\partial v} f(x,v,t)=&& \nonumber  \\ 
-\int _{-\infty }^\infty dv^\prime \, |v-v^\prime | {\bigg [  } f(x,v,t)f(x,v^\prime ,t)- \quad && \nonumber \\ 
\int _{0 }^\infty ds \, \frac {e^{ -s/\tau _{|v-v^\prime |}}}{\tau _{|v-v^\prime |}}
f(x_s,v_s^+,t-s)f(x_s,v_s ^- ,t-s){\bigg ] },   && 
\label{k:2.4} 
\end{eqnarray} 
where $x_s$ and $v_s^\pm $ are given by the right-hand side of Eq. (\ref{k:1.4}) and Eq. (\ref{k:1.5}), 
respectively, with the following substitutions: $x_c=x$, $v_{c1}=v$, $v_{c2}=v^\prime $, and $\tau =-s$. 
If we approximate the exponential kernel in the right-hand-side of Eq. (\ref{k:2.4}) by $\delta (s-\eta )$, 
$\eta \rightarrow 0+$, this right-hand-side (the 1D collisional integral) becomes zero, and we obtain 
in this limit an equation of motion of a  collisionless gas.
The Maxwell-Boltzmann distribution 
\begin{equation} 
f_0(x,v,t)=\frac {Nm\omega _\Vert }{2\pi T} \exp \left( -\frac {m\omega _\Vert ^2x^2}{2T}-
\frac {mv^2}{2T}\right) 
\label{k:2.5} 
\end{equation} 
is a stationary solution of Eq. (\ref{k:2.4}). Here we set Boltzmann's constant to 1 and thus measure 
temperature $T$ in energy units. We normalize $f_0$ in Eq. (\ref{k:2.5}) to the 
total number of atoms $N$, since the interactions are week and the mean number of compound particles 
simultaneously present in the system is much less than $N$. 

\section{Monopole mode} 
\label{l:s}
If a system state deviates from equilibrium only slightly, the typical time of return to the equilibrium 
is estimated from the linearized form of the kinetic equation \cite{Pit10}. Following the standard 
linearization scheme, we represent $f(x,v,t)=f_0(x,v,t)+f_1(x,v,t)$ where the small 
perturbation term $f_1$ describes 
the deviation from the equilibrium and is orthogonal to $f_0$: 
$\int _{-\infty }^\infty dx \int _{-\infty }^\infty dv\, f_0(x,v,t)$\, $f_1(x,v,t)=0$. 
For the sake of definiteness, we consider one specific (monopole) 1D mode \cite{Stringari2}, 
and choose the perturbation term in the form 
\begin{equation} 
f_1(x,v,t)=f_0(x,v)[\Lambda (t)(v^2- \omega _\Vert ^2x^2)+B(t)\omega _\Vert xv].
\label{k:4.1}  
\end{equation}
If we completely neglect collisions in Eq. (\ref{k:2.4}), we 
obtain the following linearized equations: 
\begin{equation} 
\dot{\Lambda }+2\omega _\Vert B=0, \qquad 
\dot{B }-2\omega _\Vert \Lambda =0. 
\label{eq:4.2} 
\end{equation} 
In the next iteration, we substitute the solutions of Eq. (\ref{eq:4.2}) 
$\Lambda (t-s)=\Lambda (t)\cos 2\omega _\Vert s +B(t)\sin 2\omega _\Vert s$ and 
$B(t-s)=-\Lambda (t) \sin 2\omega _\Vert s +B(t)\cos 2\omega _\Vert s $ 
into the right-hand-side of Eq. (\ref{k:2.4}) and finally obtain 
\begin{equation} 
\dot \Lambda +(2\omega _\Vert -\beta ) B +\gamma \Lambda =0, \quad 
\dot B -2\omega _\Vert \Lambda =0, \label{k:4.3} 
\end{equation} 
where 
\begin{equation} 
\beta =\frac {N\omega _\Vert m^3}{64 \pi T^3}\int _0^\infty du\, u^5 e^{-mu^2/(4T)} \frac {2\omega _\Vert 
\tau _u}{1+4\omega _\Vert ^2\tau _u^2} 
\label{k:4.4} 
\end{equation} 
and  
\begin{equation} 
\gamma =\frac {N\omega _\Vert m^3}{64 \pi T^3}\int _0^\infty du\, u^5 e^{-mu^2/(4T)} \frac {4\omega _\Vert ^2
\tau _u^2}{1+4\omega _\Vert ^2\tau _u^2} \, .  
\label{k:4.5} 
\end{equation}
Solution of Eqs. (\ref{k:4.5}) yields a complex eigenfrequency. Its real and imaginary parts give the oscillation 
frequency 
\begin{equation} 
\omega = 2\omega _\Vert \sqrt {1-\beta /(2\omega _\Vert ) -\gamma ^2/(16 \omega _\Vert ^2)} 
\label{k:4.6} 
\end{equation} 
(we assume weak interactions, $\beta, \, \gamma \ll \omega _\Vert $, hence, the  
oscillatios are not overdamped) and the decrement 
\begin{equation} 
\Gamma =\gamma /2 
\label{k:4.7} 
\end{equation} 
of the 1D monopole mode. 
Using the high-velocity limit Eq. (\ref{k:1.3}) we calculate the leading terms of the two relevant parameters: 
\begin{equation} 
\beta \approx \frac \nu {8\sqrt{\pi }} N\omega _\Vert , 
\label{k:4.8} 
\end{equation}
\begin{equation} 
\gamma \approx \frac {\nu ^2 \ln (1/\nu )}{12{\pi }} N\omega _\Vert ,   
\label{k:4.9} 
\end{equation}
where 
\begin{equation} 
\nu = \frac {\hbar ^2 c\omega _\Vert }{T\sqrt{mT}} .
\label{k:4.10}
\end{equation} 
The consistency of our solution requires  $\nu N\, ^<_\sim \,  1$. The decrement of the monopole mode 
is smaller than the collision-induced correction to the oscillation frequency by a factor 
$\sim \nu \ln (1/\nu )$. If we neglect $\gamma $, Eqs.~(\ref{k:4.6}, \ref{k:4.8}, \ref{k:4.10}) yield 
the known result \cite{Stringari2} for the monopole mode frequency in a 1D thermal gas. Note that 
dissipation has been introduced in \cite{Stringari2} phenomenologically, and up to now only the 
damping via essentially 3D scattering \cite{Stringari1} has been studied.  

In the strongly interacting  regime, one has to take, instead of Eq. (\ref{k:1.3}), the opposite 
($k_r\ll c$) limit of Eq. (\ref{k:1.2a}). Then $\gamma $ decreases as $c^{-4}$ as $c\rightarrow \infty $, thus 
signifying integrability restoration for a longitudinally trapped gas in the Tonks-Girardeau limit 
\cite{Minguzzi}. 
 
We consider, as an example, a gas of $N=10^3$ $^{87}$Rb atoms trapped at $T=1~\mu $K 
in a highly elongated trap with the 
radial and longitudinal trapping frequencies 
$\omega _\perp =2\pi \times 67$ kHz and $\omega _\Vert =2\pi \times 10$ 
Hz (such a strong radial trapping needed to satisfy the one-dimensionality conditions can be achieved in optical 
lattices \cite{WeissNC}). In this case $\hbar ^2 c/m=2\hbar \omega _\perp a_s$ \cite{Olsh}, 
where $a_s=5.3$~nm is the atomic 
$s$-wave scattering length, and we obtain $\nu \approx 2\times 10^{-4}$. The collisional shift of the 
monopole mode frequency is, according to Eqs. (\ref{k:4.6},~\ref{k:4.8}), approximately 0.4\%. The decrement 
of this mode is $\Gamma \approx 3\times 10^{-4}$~s$^{-1}$, which is too small to be measured experimentally. 

The method developed in the present paper applies to non-degenerate 1D gases. However, by analogy with the 
results above, we can hazard a conjecture for the estimation of the decrement of the monopole mode also in the 
case of degenerate gas. Our guess is that the relation $\Gamma \sim \nu \ln (1/\nu )\Delta \omega $ holds also in the 
latter case, but the parameter $\nu $ has to be then estimated not from Eq. (\ref{k:4.10}), but instead from the 
the collisional shift $\Delta \omega \sim \nu N \omega _\Vert $ of the monopole mode frequency (with respect to 
the frequency $2\omega _\Vert $ of this mode in an ideal gas). At maximum (in 
the Thomas-Fermi regime), $\Delta \omega \sim \omega _\Vert $ \cite{Stringari3} and, hence, $\nu \sim 1/N$.  
Then we can roughly estimate the decrement as $\Gamma \sim \ln (1/N) \omega _\Vert /N$. The time scale 
$\Gamma ^{-1}$, on which the integrability breakdown due to longitudinal trapping can be neglected, is 
long compared to a typical duration of an ultracold-atom experiment. 

\section{Conclusion} 
\label{ccc}
We find that the combination of two effects, (i) finite time delay in wave packet propagation 
associated with a 1D collision and 
(ii) longitudinal trapping lift the integrability of a system of identical bosons in 1D with contact 
interactions. To quantify the effect of non-integrability for a non-degenerate gas, we derive a set of 
kinetic equations and apply them to the analysis of the monopole mode of a 1D gas. We find this 
decrement to be small enough to be neglected in experiment and state therefore that the main source of 
non-integrability and thermalization in cold atomic Bose-gases in elongated traps in the 1D regime is 
virtual excitation of transversal degrees of freedom that gives rise to effective velocity-changing 
3-body elastic collisions \cite{Mazets1,Mazets2,Glazman}. 

The author thanks J. Schmiedmayer and H.-C. N\"agerl for helpful discussions. This work is supported 
by the FWF (project P22590-N16).

\end{document}